\documentclass[letterpaper, 10 pt, conference]{ieeeconf}

\IEEEoverridecommandlockouts                              

\usepackage{amsmath} 

\usepackage{multirow}
\usepackage{multicol}
\usepackage{rotating}
\usepackage{subfigure}
\usepackage{algorithm}

\usepackage{algorithmic}

\usepackage{cleveref}
\usepackage{xcolor}
\DeclareMathOperator*{\argmax}{arg\,max}
\DeclareMathOperator*{\argmin}{arg\,min}

\newcommand{\updated}[1]{#1}
\newcommand{\finalUpdated}[1]{\textcolor{black}{#1}}

\newcommand{\ignore}[1]{}

\begin{document}

\title{\LARGE \bf
\updated{Integrated Task Assignment and Path Planning} for \\ Capacitated Multi-Agent Pickup and Delivery
}

\author{Zhe Chen$^1$, Javier Alonso-Mora$^2$, Xiaoshan Bai$^3$, Daniel D. Harabor$^1$, and Peter J. Stuckey$^1$ %
\thanks{This research was supported in part by the Australian Research Council under grants DP190100013 and DP200100025, by Amazon Research Awards, by the European Union’s Horizon 2020 research and innovation programme under grant 101017008, and by the National Natural Science Foundation of China under grant 62003217. ($Corresponding~ author: Xiaoshan~ Bai$.)} 

\thanks{$^1$Zhe Chen, Daniel D. Harabor and Peter J. Stuckey are with the Faculty of Information Technology, Monash University, Wellington Rd, Clayton VIC 3800, Australia (e-mail:{\tt\footnotesize \{zhe.chen, daniel.harabor, peter.stuckey\}@monash.edu}).}
\thanks{$^2$Javier Alonso-Mora is with the Dep. of Cognitive Robotics, Delft University of Technology, Delft 2628 CD, The Netherlands (e-mail: {\tt\footnotesize j.alonsomora@tudelft.nl}).}
\thanks{$^3$Xiaoshan Bai is with the College of Mechatronics and Control Engineering, Shenzhen University, Shenzhen 518060, China, and also with the Dep. of Cognitive Robotics, Delft University of Technology, Delft 2628 CD, The Netherlands (e-mail: {\tt\footnotesize baixiaoshan@szu.edu.cn}).}
}


\maketitle

\begin{abstract}
Multi-agent Pickup and Delivery (MAPD) is a challenging industrial problem where a team of robots is tasked with transporting a set of tasks, 
each from an initial location and each to a specified target location. Appearing in the context of automated warehouse logistics and automated mail sortation, MAPD
requires first deciding which robot is assigned what task (i.e., Task Assignment or TA) followed by a subsequent coordination problem where each robot must
be assigned collision-free paths so as to successfully complete its assignment (i.e., Multi-Agent Path Finding or MAPF).
Leading methods in this area solve MAPD sequentially: first assigning tasks, then assigning paths. In this work we propose a new 
coupled method where task assignment choices are informed by actual delivery costs instead of by lower-bound estimates. 
The main ingredients of our approach are a marginal-cost assignment heuristic and a meta-heuristic improvement strategy based on Large Neighbourhood Search.
As a further contribution, we also consider a variant of the MAPD problem where each robot can carry multiple tasks instead of
just one. 
Numerical simulations show that our approach yields efficient and timely solutions and we report significant improvement compared with other recent methods from the literature.
\end{abstract}


\section{Introduction} 
\label{intro}




  
\begin{figure}[t!]
    \subfigure[\label{fig:amazon}]{\includegraphics[height=6em,width=11.3em]{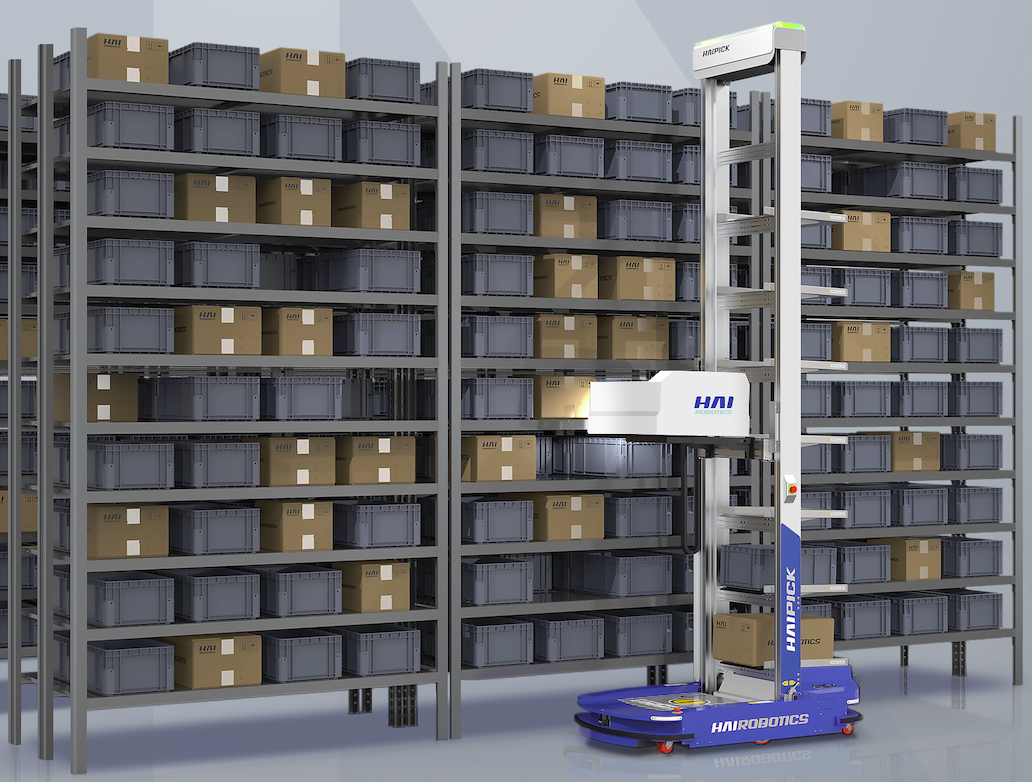}}
    \hfill
    \subfigure[\label{fig:alibaba}]{\includegraphics[height=6em,width=11.3em]{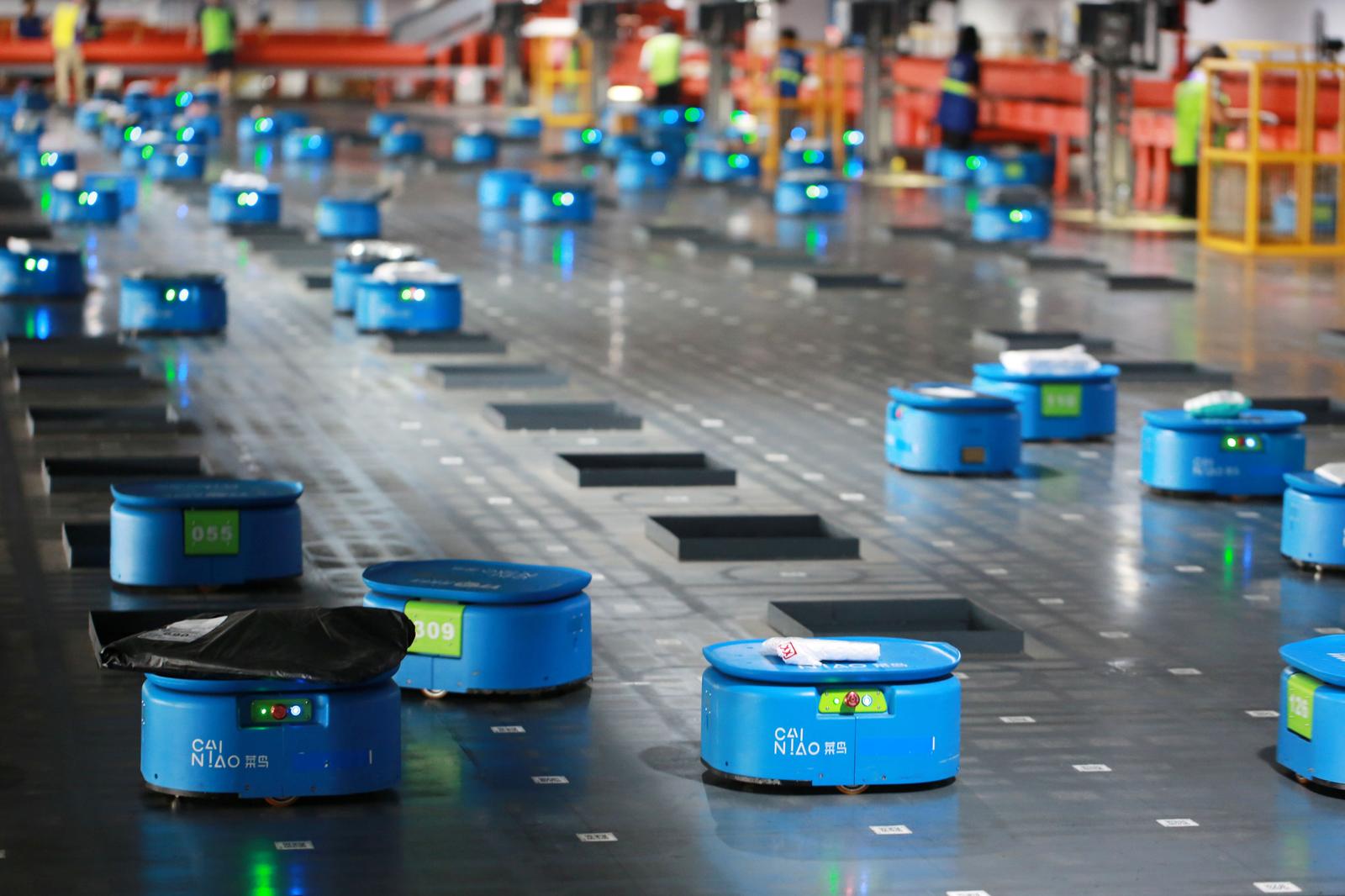}}
    \hfill
    \label{fig:applications}
    \caption{ MAPD applications: (a)  \updated{an automated fulfillment center with robots carrying multiple objects~\cite{hai_robots}}; (b) an automated sortation centre~\cite{meng2019idle}.
    }
\end{figure}

In automated warehouse systems, a team of robots works together to fulfill  a set of customer orders. Each order comprises one or more items found on the warehouse floor, which must be
delivered to a picking station for consolidation and delivery. In automated sortation centres, meanwhile, a similar problem arises. Here, the robotic team is tasked with carrying mail tasks
from one of several emitter stations, where new parcels arrive, to a bin of sorted tasks, all bound for the same processing facility where they will be dispatched for delivery. 
Illustrated in Fig. \ref{fig:applications}, such systems are at the heart of logistics operations for major online retailers such as Amazon and Alibaba.
Practical success in both of these contexts depends on computing timely solutions to a challenging optimization problem known in the literature as 
Multi-agent Pickup and Delivery (MAPD)~\cite{ma2017lifelong}.  

In MAPD, we are given a set of tasks (equiv. packages) and a team of cooperative agents (equiv. robots).
Our job is twofold: first, we must assign every task to some robot; second, we need to find for each robot
a set of collision-free paths that guarantee every assigned task to be successfully completed. 
Each of these aspects (resp. Multi-robot task assignment (TA)~\cite{korsah2013comprehensive} and Multi-agent Path Finding (MAPF)~\cite{SternSoCS19}) 
is itself intractable, which makes MAPD extremely challenging to solve in practice. 
Further complicating the situation is that the problem is {\em lifelong} or {\em online}, 
which means new tasks arrive continuously and the complete set of tasks is a priori unknown.

A variety of different approaches for MAPD appear in the recent literature. 
Optimal algorithms, such as CBS-TA~\cite{honig2018conflict}, guarantee solution quality 
but at the cost of scalability: only small instances can be solved and timeout failures 
are common. 
Decentralised solvers, such as TPTS~\cite{ma2017lifelong}, can scale to problems with 
hundreds of agents and hundreds of tasks but at the cost of solution quality:
assignments are greedy and made with little regard to their impact on
overall solution costs.
Other leading methods, such as TA-Hybrid~\cite{liu2019task}, suggest a middle road:
MAPD is solved centrally but as a sequential two-stage problem: task assignment first
followed by coordinated planning after. The main drawback in this case is that the
assignment choices are informed only by lower-bound delivery estimates instead of actual costs. In other words, the cost of the path planning task may be far higher
than anticipated by the task assignment solver.

In this work we consider an alternative approach to MAPD which solves task 
assignment and path planning together. 
\updated{We design a marginal-cost assignment heuristic and a meta-heuristic improvement strategy}  to 
match tasks to robots. The costs of these assignments are evaluated by 
solving the associated coordination problem using prioritised 
planning~\cite{silver2005cooperative}.
We then iteratively explore the space of possible assignments by 
destroying and repairing an incumbent solution using Large 
Neighbourhood Search~\cite{LNS}.
We give a complete description of this algorithm and we report convincing
improvement in a range of numerical simulations vs. \updated{the Token Pass and Task Swap (TPTS) algorithm in  \cite{ma2017lifelong}}, 
arguably the current state-of-the-art sub-optimal method in this area. 
As a further contribution we also consider and evaluate a natural
extension of the MAPD problem where each agent is allowed to 
carry more than one task at a time, reflecting emerging robotic warehouse systems (see e.g.~\cite{hai_robots}, \Cref{fig:applications} (a)). For comparison, all other 
work in the literature assume the capacity of each agent is always 
$1$ which implies immediate delivery is required after every pickup. 
We show that in the generalised case solution costs can decrease
substantially, allowing higher system performance with the same
number of agents.

\section{Related Work} 

\subsection{Task Assignment}
 
The problem studied in this paper requires both the task assignment of robots and the planning of collision-free paths. Nguyen $et~al.$~\cite{nguyen2019generalized} solved a generalised target assignment and path finding problem with answer set programming. They designed an approach operating in three phases for a simplified warehouse variant, where the number of robots is no smaller than the number of tasks and unnecessary waiting of agents exists between the three phases. As a result, the designed approach scales only to $20$ tasks or robots. 

The task assignment aspect of the studied problem is related to multi-robot task allocation problems, which have been widely
studied \cite{korsah2013comprehensive,bai2020event}. Most closely related are the VRP \cite{toth2014vehicle} and its variants \cite{bai2018integrated}, all of which are NP-hard problems. The pickup and delivery task assignment problems have also received attention \cite{bai2019efficient,cordeau2007dial}. 
In \cite{bai2019efficient}, the package delivery task
assignment for a truck and a drone to serve a set of customers with precedence constraints was investigated, where several heuristic assignment algorithms are proposed. 
Cordeau and Laporte \cite{cordeau2007dial} conducted a review on the dial-a-ride problem, where the pickup and delivery requests for a fleet of vehicles to transport a set of customers need to respect the customers’ origins and destinations. 
\updated{In \cite{tillman1972upperbound}, the original concept of regret for not making an assignment may be found to assign customers to multiple depots in a capacity-constrained routing, where the regret is the absolute difference between the best and the second best alternative.  
For the vehicle routing and scheduling
problem with time windows in \cite{potvin1993parallel}, Potvin and Rousseaua used the sum of the
differences between the best alternative and all the other alternatives as the regret to route each customer. Later on, in \cite{koenig2008agent}, agent coordination with regret clearing was studied. In the paper, each task is assigned to the agent whose regret is largest, where the regret of the task is the difference between the defined team costs resulting from assigning the task to the second best and the best agent. }
But all the methods above avoid reasoning about collisions of vehicles, they assume, quite correctly for vehicle routing, that routes 
of different vehicles do not interfere.  This assumption does not hold however for
automated warehouses or sortation centres.

\subsection{Multi-agent Pickup and Delivery}

For warehouses or sortation centres, it is necessary to consider the interaction between agent routes. The MAPD problem describes this scenario. Ma~\emph{et al}~\cite{ma2017lifelong} solves the MAPD problem online in decentralised manner using a method similar to \emph{Cooperative A*} \cite{silver2005cooperative}, 
and in a centralised manner, which first greedily assigns tasks to agents using a Hungarian Method and then uses \emph{Conflict Based Search} (CBS) \cite{sharon2015conflict} to plan collision-free paths. Liu~\emph{et al}~\cite{liu2019task} proposed TA-Hybrid to solve the problem offline, which assumes all incoming tasks are \updated{known initially}. TA-Hybrid first formulates the task assignment as a travelling salesman problem (TSP) and solves it using an existing TSP solver. Then it \updated{plans} collision-free paths using a CBS-based algorithm. 

Researchers have also investigated how to solve this problem optimally. 
Honig~\emph{et al}~\cite{honig2018conflict} proposed CBS-TA, which solves the problem optimally by modifying CBS to search an assignment search tree. 
However, solving this problem optimally is challenging, which leads to the poor scalability of CBS-TA. Other limitations of CBS-TA and \emph{TA-Hybrid} are that they are both offline 
and hard to adapt to work online, and they don't allow an agent to carry multiple items simultaneously.  
   
\subsection{Multi-agent Path Finding}
    
Multi-agent path finding (MAPF) is an important part of MAPD problem and is well studied. Existing approaches to solve MAPF problems are categorised as optimal solvers, bounded-suboptimal solvers, prioritised solvers, rule-based solvers, and so on. Optimal solvers include \emph{Conflict Based Search} (CBS) \cite{sharon2015conflict}, \emph{Branch-and-Cut-and-Price} (BCP) \cite{bettinelli2011branch}, A* based solvers \cite{EPEA} and Reduction Based Solvers \cite{SurynekIJCAI19}. These solvers solve the problem optimally and their weakness is the poor scalability. Bounded-suboptimal solvers such as \emph{Enhanced CBS} (ECBS) \cite{CohenIJCAI16} can scale to larger problems to find near optimal solutions. Prioritised solvers plan paths for each agent individually and avoid collisions with higher priority agents. The priority order can be determined before planning as in \emph{Cooperative A*} (CA) \cite{silver2005cooperative}, or determined on the fly as in \emph{Priority Based Search} (PBS) \cite{PBS}. Rule-base solvers like 
\emph{Parallel Push and Swap} \cite{PLPushAndSwap} 
guarantee to find solutions to MAPF in polynomial time, 
but the quality of these solutions is far from optimal.
Some researchers focus on the scalability of online multi-agent path finding in MAPD problem. \emph{Windowed-PBS}~\cite{li2020lifelong} plans paths for hundreds of agents in MAPD problem,
however it assumes that tasks are assigned by another system. 

\subsection{Practical Considerations}
This research focuses on the task assignment and path planning for real world applications.  However, it also needs to consider plan execution and kinematic constraints necessary to achieve a computed plan in practice. 

One issue that can arise in practice is unexpected delays, such as those that can be caused by a robot's mechanical differences, malfunctions, or other similar issues. \updated{Several robust plan execution policies were designed in \cite{honig2019} and \cite{ma2017multi}} to handle unexpected delays during execution. The plans generated by our algorithms can be directly and immediately combined with these policies. Furthermore, \updated{$k$-robust planning was proposed in \cite{atzmon}}, which builds robustness guarantees into the plan. Here an agent can be delayed by up to $k$ timesteps and the plan remains valid. Our algorithms can also adapt this approach to generate a $k$-robust plan. 

Actual robots are further subject to kinematic constraints, which are not considered by our MAPF solver. To overcome this issue, \updated{a method was introduced in \cite{honig2016multi}} for post-processing a MAPF plan to derive a plan-execution schedule that considers a robot's maximum rotational velocities and other properties. This approach is compatible with and applicable to any MAPF plan computed by our approach.


\section{Problem formulation}
\label{Pro}
Consider that multiple dispersed robots need to transport a set of tasks from their initial dispersed workstations to corresponding destinations while avoiding collisions, where each task has a release time, that is the earliest time to be picked up. The robots have a limited loading capacity, which constrains the number of tasks that each robot can carry simultaneously. Each robot moves with a constant speed for transporting the tasks and stops moving after finishing its tasks. 
The objective is to minimise the robots' total travel delay (TTD) to transport all the tasks while avoiding collisions.

\subsection{Formula Definition As An Optimisation Problem}

We use $\mathcal P=\{1,\cdots,n\}$ to denote the set of indices of $n$ randomly distributed tasks that need to be transported from their initial locations to corresponding dispersed destinations. 
Each task $i \in \mathcal P$ is associated with a given tuple $(s_i,g_i,r_i)$, where $s_i$ is the origin of $i$, $g_i$ is the destination of $i$, and $r_i$ is the release time of $i$. 
$\mathcal R=\{n+1,\cdots,n+m\}$ denotes the set of indices of $m>1$ robots that are initially located at dispersed depots. We use $s_k$ to represent the origin of robot $k \in \mathcal R$. 
To transport task $i$, one robot needs to first move to the origin $s_i$ of $i$ to pick up the task no earlier than its release time $r_i$, and then transport the task to its destination $g_i$.  
It is assumed that the robots can carry a maximum of $C$ tasks at any time instant. 
Let $n_k(t) \leq C$ be the number of tasks carried by robot $k\in \mathcal R$ at time instant $t$, and $p_k(t)$ be the position of robot $k$ at $t$. \updated{We model the operation environment as a graph consisting of evenly distributed vertices and edges connecting the vertices, and assume that the tasks and robots are initially randomly located at the vertices.
When the robots move along the edges in the graph, they need to avoid collision with each other: so two robots cannot be in the same vertex at the same time instant $t$, and they also cannot move along the same edge in opposite directions at the same time.}
Let $\mathcal I=\{s_1,...,s_{n+m},g_1,...,g_n\}$, and   
$t(i,j)$ denote the shortest time for a robot to travel from $i$ to $j$ for each pair of $i,j \in \mathcal I$. Trivially, $t(i,i)=0$ for each $i \in \mathcal I$.

Let $\sigma_{ijk}: \mathcal I \times \mathcal I \times \mathcal R \rightarrow \{0,1\}$ be the path-planning mapping that maps the indices $i,j \in \mathcal I$ of the starting and ending locations and $k\in \mathcal R$ of the $k$th robot to a binary value, which equals one if and only if it is planned that robot $k$ directly travels from location $i$ to location $j$ \updated{for performing a pick-up or drop-off operation for transporting the tasks associated with the locations}. So $\sigma_{iik} = 0$ for all $i\in \mathcal I$ and $k\in \mathcal R$. 
Let the task-assignment mapping $\mu_{ik}: \mathcal P \times \mathcal R \rightarrow \{0,1\}$ map the indices $i\in \mathcal P$ of the $i$th task and $k\in \mathcal R$ of the $k$th robot to a binary value, which equals one if and only if it is planned that robot $k$ picks up task $i$ at $s_i$ no earlier than $r_i$ and then transports $i$ to its destination. 
We use variable $a(j)$, initialised as $a(j)=0$, to denote the time when a robot \updated{performs a pick-up or drop-off operation at location $ j\in \mathcal I$ to transport a task. Thus, $n_k(a(s_i)+1) = n_k(a(s_i))+1$ if $p_k(a(s_i))=s_i$, and $n_k(a(g_i)+1) = n_k(a(g_i))-1$  if $p_k(a(g_i))=g_i$, $\forall i\in \mathcal P, \forall k\in \mathcal R$. }

Then, the objective to minimize the total travel delay (TTD) for the robots to  transport all the tasks while avoiding collisions is to minimise 
\begin{eqnarray}\label{eq:f}
f=\sum_{i \in \mathcal P} (a(g_i)-(r_i+t(s_i,g_i))), 
\end{eqnarray}
subject to  
\begin{align}
\sum_{j\in \mathcal I} \sigma_{js_ik}  & = \sum_{j\in \mathcal I} \sigma_{s_ijk}, ~ \forall i\in \mathcal P,  \forall k\in \mathcal R;
\label{eq:1}\\
\sum_{j\in \mathcal I} \sigma_{js_ik}  & =  \mu_{ik}, ~ \forall i\in \mathcal P,  \forall k\in \mathcal R; \label{eq:2}\\
\sum_{k\in \mathcal R} \mu_{ik} & =  1,~ \forall i\in \mathcal P; \label{eq:3}\\
\sigma_{ijk}\updated{\cdot} (p_k(a(i))-i)  & =  0, ~ \forall i,j\in \mathcal I,  \forall k\in \mathcal R; \label{eq:4}\\
\sigma_{ijk}\updated{\cdot} (p_k(a(j))-j)  & =  0, ~ \forall i,j\in \mathcal I,  \forall k\in \mathcal R; \label{eq:5}\\
r_i & \leq  a(s_i),~ \forall i\in \mathcal P; \label{eq:6}\\
\sigma_{ijk}\updated{\cdot}(a(i)+t(i,j))& \leq  a(j),~ \forall i,j\in I,  \forall k\in \mathcal R; \label{eq:7} \\
n_k(t)  & \leq  C, \forall k\in \mathcal R, \forall t; \label{eq:8}\\
p_k(t)& \neq   p_w(t),~ \forall k,w \in \mathcal R, \updated{k\neq w}, \forall t; \label{eq:9} \\
(p_k(t),p_k(t+1))& \neq   (p_w(t+1),p_w(t)),~ \forall k,w \in \mathcal R, \forall t;~~~~ \label{eq:10} ~\\
\sigma_{ijk}, \mu_{ik} &\in  \{0,1\}, \forall i,j\in \mathcal I,  \forall k\in \mathcal R. \label{eq:11} \nonumber
\end{align}
Constraint (\ref{eq:1}) requires that \updated{the same robot drops off the task picked up by it};  (\ref{eq:2}) denotes that \updated{a task will be transported by a robot if the robot picks up the task}; 
(\ref{eq:3}) implies that each task is transported by exactly one robot;
(\ref{eq:4}) and (\ref{eq:5}) require that vehicle $k$ will visit all the locations, planned to be visited, at certain time instants; 
(\ref{eq:6}) guarantees
that the earliest time for the robots to pickup every task is the time when the task is released; 
(\ref{eq:7}) ensures
that there
is no \updated{shorter time for each robot to
move between  two  arbitrary locations $i$ and $j$ compared with $t(i,j)$}; 
(\ref{eq:8}) guarantees that the robots' capacity constraint is always satisfied;  (\ref{eq:9}) and (\ref{eq:10}) require that there is no collision between any two robots.

\section{Task Assignment and Path Planning} 

Existing MAPD algorithms perform task assignment and path planning separately. \updated{Here we propose several algorithms for simultaneous task assignment and path planning}, and path costs from planning are used to support the task assignment. 

\subsection{Task Assignment Framework}
\label{sec:ta-framework}

\begin{figure}[t]
\centering
\includegraphics[width=1\columnwidth]{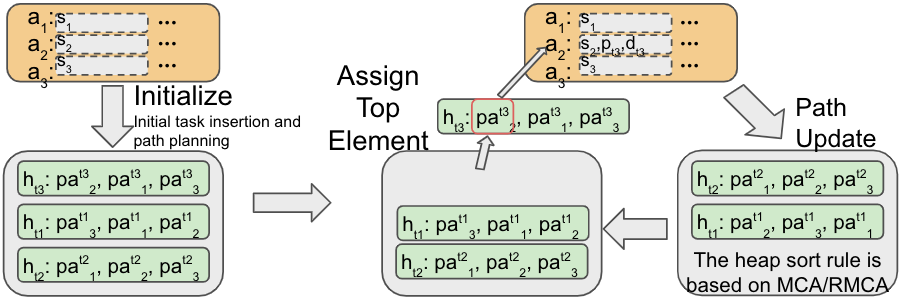}

\caption{\updated{The flowchart of MCA/RMCA for assigning three tasks/packages $\{t_1,t_2,t_3\}$ to three robots $\{1,2,3\}$. The gray box is priority heap $\mathcal{H}$, green box is potential assignment heap $h$, orange box is current assignment set $\mathcal{A}$, dashed border box is ordered action sequence $o_i$ for each robot $i$, $s_i$ is $i$'s initial location, and $p_{t3}$ and $d_{t3}$ are respectively the pick-up and destination location of task $t_3$.} }
\label{figure:flow}
\end{figure}

\begin{algorithm}[!tp]
	\small 
	\caption{Simultaneous Task Assignment and Path Planning}
	\label{algor:TA}
	\begin{algorithmic}[1]
		\REQUIRE Current Assignment Set $\mathcal A$, task set $\mathcal P$, robot set $\mathcal R$, and the loading capacity $C$.
		
		\STATE $\mathcal P^u \gets \mathcal P$
		
		\STATE $\mathcal H \gets$ build potential assignment heaps based on $\mathcal A$
		\WHILE {$\mathcal P^u \neq \emptyset$}
		    \STATE $pa^i_k \gets \mathcal H.top().top()$
		    \STATE $\mathcal{A} \gets (\mathcal{A} - \{a_k\}) \cup \{pa^i_k\}$ 
    		\STATE $a_k \gets pa^i_k$
    		\STATE Delete $i$ from $\mathcal P^u$
    		\STATE Delete $h_{i}$ from $\mathcal H$
    		\FOR{$h_{j} \in \mathcal H$}
    		    \STATE // Update $pa^{j}_k$ based on $a_k.o_k$
    		    \STATE $pa^{j}_k \gets$ Get assignment of $j$ on $k$ from $h_{j}$ 
    		    \STATE $pa^{j}_k.o^{j}_k \gets insert(j,a_k.o_k)$
    		    \STATE $pa^{j}_k.path \gets planPath(pa^{j}_k.o^{j}_k)$
    		    \STATE $h_j.update(pa^{j}_k)$
    		    \STATE // Update top elements' paths
    		    \STATE $updateHeapTop(h_{j}, a_k, 1+(RMCA))$ // Algorithm 2
    		\ENDFOR
		\ENDWHILE
		
		\RETURN $\mathcal A$
		
		
	\end{algorithmic} 
\end{algorithm}

Fig. \ref{figure:flow} shows the overall process of how task assignment and path planning are performed simultaneously. The key component of this approach is a current assignment set $\mathcal A$ and a priority heap $\mathcal H$. 
$\mathcal A$ stores a set of assignments $a_k$ \updated{which contains $o_k$, an ordered sequence of actions (pick-up and drop-off each task) assigned to each robot $k \in \mathcal R$,  $k$'s current collision-free $path$, and the TTD for $k$ to transport the assigned tasks}. \updated{$o_k$ is initialized as $\{s_k\}$, and $t(o_k)$ is used to denote the TTD for robot $k$ to transport all the tasks by following $o_k$. }
The priority heap $\mathcal H$ stores a set of potential assignment heaps $h_i$, one for each unassigned task $i \in \mathcal P$. A potential assignment heap $h_i$  for task $i$ stores all potential assignments of $i$ to each robot $k \in \mathcal R$ based on $k$'s current assignment $a_k$. 
An entry in the heap $h_i$ is a potential assignment $pa^i_k$ of task $i$ to robot $k$ which
includes updated versions of $o_k$ and a revised path and cost for the agent under the addition of task $i$ to robot $k$. \updated{The algorithm continues assigning tasks from the unassigned task set $\mathcal P^u$ initialized as $\mathcal P$}, and keeps updating $\mathcal H$ until all tasks are assigned.

\Cref{algor:TA} shows the pseudo-code for task assignment framework.
At the start of the algorithm, $\mathcal A$ has no assigned tasks and paths. $\mathcal H$ is initialized to include one potential assignment heap for each task. Each potential assignment heap tries to assign the task $i$ to every robot based on $\mathcal A$. 

The main while loop of the algorithm keeps selecting and assigning the top potential assignment $pa^i_k$ of the top potential assignment heap of $\mathcal H$. The potential assignment $pa^i_k$ assigns task $i$ to robot $k$. Then the $a_k \in \mathcal A$ is replaced by $pa^i_k$, $h_i$ is deleted from $\mathcal H$ and $i$ deleted from $\mathcal P^u$. When the action sequences $o_k$ and path for robot $k$ in $\mathcal A$ change, all other potential assignment's action sequence $o^{j}_k$ on robot $k$ in any $h_{j}, j \in \mathcal P^u/\{i\}$, must be recalculated based on the new path for agent $k$. 

The behaviour of $insert()$ function in Algorithm \ref{algor:TA} will be explained in section \ref{sec:mca} and section \ref{sec:rmca}. 
The $planPath()$ function uses prioritised planning with space-time A* \cite{silver2005cooperative}, which is fast and effective, to plan a single path for agent $k$ following its ordered action sequence $o_k$ while avoiding collisions with any other agents' existing paths in $\mathcal A$. As a result, the overall priority order for path planning is decided by the task assignment sequence. \finalUpdated{It is worth noting that the path planning part of Algorithm $1$ might be incomplete as the prioritised planning is known to be incomplete~\cite{PBS}.}

For the remaining potential assignments on robot $k', k' \neq k, k' \in \mathcal R$ in any $h_{j}$, the recalculation of action sequence $o^{j}_{k'}$ is not necessary since the assigned tasks $a_{k'} \in \mathcal A$ do not change. 
However their current paths may collide with the updated agents path $a_k.path$. 
To address this issue, we could check for collisions of all potential assignments for agents other than  $k$ and update their paths if they collide with the new path for agent $k$.
A faster method is to only check and update the paths for assignments at the top $v$ elements of each potential 
assignment heap using the $updateHeapTop()$ function \updated{shown in Algorithm \ref{algor:topMCA}}. 
Using the second method saves \updated{considerable} time and it only slightly influences the task assignment outcome.

A potential assignment heap sorts each potential assignment 
in increasing order of marginal cost. The sorting order of $\mathcal H$ is decided by the task selection methods defined below.
\ignore{
In section \ref{sec:mca} and section \ref{sec:rmca}, we introduced two task selection methods. The $updateHeapTop()$ function also differs for different task selection methods and their behaviours will be introduced in the next two sections. 
}

\begin{algorithm}[!tp]
	\small 
	\caption{Update potential assignment heap for (R)MCA}
	\label{algor:topMCA}
	\begin{algorithmic}[1]
		\REQUIRE Assignment heap $h_{j}$, new assignment $a_k$, limit $v$
	    \WHILE{$\exists$ element $pa^j_l$ in top $v$ elements of $h_{j}$ with collision with $a_k.path$}
	        \STATE $pa^j_l.path \gets planPath(pa^{j}_{l}.o^j_{l}, a_k)$
	        \STATE $h_{j}.updateTop(v)$
	    \ENDWHILE
		
	\end{algorithmic} 
\end{algorithm}

\ignore{
\begin{algorithm}[!tp]
    \label{algor:topMCA}

	\small 
	\caption{
        Updating potential assignment heap for MCA }

	\begin{algorithmic}[1]
		\REQUIRE Potential assignment heap $h_{i'}$, current assignment $a_k$
		
	    \WHILE{$h_{i'}.top()$ has collision with $a_k.path$}
	        \STATE $h_{i'}.top().path \gets planPath(pa^{i'}_{top}.o^{i'}_{top}, a_k)$
	        \STATE $h_{i'}.updateTop()$
	    \ENDWHILE
		
	\end{algorithmic} 
\end{algorithm}

\begin{algorithm}[!tp]
\label{algor:RMCAtop}
	\small 
	\caption{
        Updating potential assignment heap for RMCA }
	\begin{algorithmic}[1]
		\REQUIRE Potential assignment heap $h_{i'}$, current assignment $a_k$
		
	    \WHILE{$h_{i'}.top()$ has collision with $a_k.path$ \OR $h_{i'}.second()$ has collision with $a_k.path$}
	        \STATE $ h_{i'}.top().path \gets planPath(h_{i'}.top().o^{i'}_{top},a_k)$
	        \STATE $ h_{i'}.second().path \gets planPath(h_{i'}.second().o^{i'}_{second},a_k)$
	        \STATE $h_{i'}.updateTopTwo()$

	    \ENDWHILE
		
	\end{algorithmic} 
\end{algorithm}
}

\subsection{Marginal-cost Based Task Selection} 
\label{sec:mca}

We now introduce the marginal-cost based task \updated{assignment algorithm} (MCA).  
The target of MCA is to select a task $i^\star$ in $\mathcal P^u$ to be assigned to robot $k^\star \in \mathcal R$, with action sequences $q^\star_1$ and $q^\star_2$ for $k^\star$ to pick up and deliver $i^\star$, while \updated{satisfying}: 
\begin{equation}
(k^\star,i^\star,q^\star_1,q^\star_2) = \mathop{\argmin}_{\substack{k \in \mathcal R, i\in \mathcal P^u,\\ 1<q_1\leq |o_k|,  \\q_1<q_2\leq |o_k|+1}} \{t((o_k\oplus_{q_1}s_i)\oplus_{q_2}g_i) - t(o_k)\},
\label{eq:MTAA1}
\end{equation}
\noindent 
where operator $(o_k\oplus_{q_1}s_i)\oplus_{q_2}g_i$ means to first insert location $s_i$ at the $q_1$th position of the current route $o_k$, and then insert location $g_i$ at the $q_2$th position of the current $o_k$. If $q_1= |{o_k}|$, $s_i$ is inserted to the second last of $o_k$ where $|{o_k}|$ is the length of ${o_k}$ and the last action should always be go back to start location. 
\updated{After assigning task $i^\star$ to robot $k^\star \in \mathcal R$, the unassigned task set $\mathcal P^u$ is updated to $\mathcal P^u=\mathcal P^u \setminus \{i^\star\}$, and $k^\star$'s route is updated to $o_{k^\star}= (o_{k^\star} \oplus_{q^\star_1} s_{i^\star}) \oplus_{q^\star_2} g_{i^\star}$.} 

To satisfy equation (\ref{eq:MTAA1}), the $insert()$ function in \updated{Algorithm} \ref{algor:TA} tries all possible combinations of $q^\star_1$ and $q^\star_2$ and selects $q^\star_1$ and $q^\star_2$ that \updated{minimise the incurred marginal TTD by following $o_k$  while ignoring collisions for transporting task $i^\star$,} 
where $k$'s load is always smaller than capacity limit $C$. Then the $planPath()$ function \updated{uses} an $A^\star$ algorithm to plan a path following $o^{i}_k$, while avoiding collision with any $a_{k'}.path, a_{k'}\in \mathcal A, k' \neq k$, and calculates the real marginal cost in terms of TTD. Finally, the $updateHeapTop()$ function (Algorithm \ref{algor:topMCA} with $v = 1$) updates the potential assignment heaps. 
The heap of potential assignment heaps $\mathcal H$ sorts potential assignment heaps based on marginal cost of the top potential assignment $pa^i_{top}$ of each potential assignment heap $h_i$ in increasing order, where $i \in \mathcal P^u$.

\subsection{Regret-based Task Selection}
\label{sec:rmca}
This section introduces a regret-based MCA (RMCA), which incorporates a form of look-ahead information to select the proper task to be assigned at each iteration. 
\updated{Inspired by \cite{tillman1972upperbound,koenig2008agent}, RMCA chooses the next task to be assigned} based on the difference in the marginal
cost of inserting the task into the best robot's route and the
second-best robot's route, \updated{and then assigns the task to the robot that has
the lowest marginal cost to transport the task.} 

For each task $i$ in the current unassigned task set $\mathcal P^u$, we use $k^*_1$ to denote the robot that inserting $i$ into its current route \updated{with the smallest incurred} marginal travel cost while avoiding collisions, where 
\begin{equation}\label{eq:RMCA1}
(k^\star_1,q^\star_1,q^\star_2)  = \mathop{\argmin}_{\substack{k_1 \in \mathcal R, \\ 1<q_1\leq |o_k|,  \\q_1<q_2\leq |o_k|+1}} \{t((o_k\oplus_{q_1}s_i)\oplus_{q_2}g_i) \\ - t(o_k)\}.
\end{equation}
The second-best robot  $k^*_2 \in \mathcal R \setminus \{k^*_1\}$ to serve $i$ is 
\begin{equation}\label{eq:RMCA2}
(k^\star_2,p^\star_1,p^\star_2)  = \mathop{\argmin}_{\substack{k_2 \in \mathcal R \setminus \{k^*_1\}, \\ 1<p_1\leq |o_k|,  \\p_1<p_2\leq |o_k|+1}} \{t((o_k\oplus_{p_1}s_i)\oplus_{p_2}g_i) \\ - t(o_k)\}.
\end{equation}

Then, we propose two methods for RMCA to determine which task $i^*\in \mathcal P^u$ will be assigned. 

The first method, \textbf{RMCA(a)}, uses \emph{absolute regret} which is commonly used in other regret-based algorithms. The task selection satisfies: 
\begin{equation}\label{eq:RMCA3}
i^\star  = \mathop{\argmax}_{i \in \mathcal P^u} ~ t((o_{k^\star_2}\oplus_{p^\star_1}s_i)\oplus_{p^\star_2}g_i)\\- t((o_{k^\star_1}\oplus_{q^\star_1}s_i)\oplus_{q^\star_2}g_i).
\end{equation}   

The second method, \textbf{RMCA(r)}, uses \emph{relative regret} \updated{to select} a task satisfying the following equation:
\begin{equation}\label{eq:RMCA4}
i^\star  = \mathop{\argmax}_{i \in \mathcal P^u} ~ t((o_{k^\star_2}\oplus_{p^\star_1}s_i)\oplus_{p^\star_2}g_i)\\/ t((o_{k^\star_1}\oplus_{q^\star_1}s_i)\oplus_{q^\star_2}g_i).
\end{equation}

Both RMCA(r) and RMCA(a) use the same $insert()$ function in section \ref{sec:mca} to select an insert location for each potential assignment. The main difference between RMCA and MCA is that the heap $\mathcal H$ sorts the potential assignment heaps $h_{i}, i \in \mathcal P^u$ by absolute or relative regret.
\ignore{
$h_{i}.second().cost/h_{i}.top().cost$, for RMCA relative, or by $h_{i}.second().cost-h_{i}.top().cost$, for RMCA absolute.} 
RMCA uses Algorithm \ref{algor:topMCA} with $v = 2$ to 
ensure that the top two elements of each heap are kept up to date.

\subsection{Anytime Improvement Strategies}

\begin{algorithm}[!tp]
	\small 
	\caption{Anytime Improvement Strategy}
	\label{algor:anytime}
	\begin{algorithmic}[1]
		\REQUIRE A set of current assignment $\mathcal A$, Group size $n$, $time\ limit$
		
	    \WHILE{$runtime < time\ limit$}
	        \STATE $\mathcal A',P^u \gets destroyTasks(\mathcal A, n)$
	        \STATE $\mathcal A' \gets RMCA(\mathcal A',\mathcal P^u)$
	        \IF{$\mathcal A'.cost \leq \mathcal A.cost$}
	            \STATE $\mathcal A = \mathcal A'$
	        \ENDIF
	        
	    \ENDWHILE
	    \RETURN A set of current assignment $\mathcal A$
		
	\end{algorithmic} 
\end{algorithm}

After finding an initial solution based on RMCA, we make use of an anytime improvement strategy on the solution. This strategy is based on the concept of Large Neighbourhood Search (LNS)~\cite{LNS}. As shown in Algorithm \ref{algor:anytime}, the algorithm will continuously destroy some assigned tasks from the current solution and reassign these tasks using RMCA. If a better solution is found, we adopt the new solution,
and otherwise we keep the current solution.
We keep destroying and re-assigning until time out. We propose three neighbour selection strategies to select tasks to destroy. 

\subsubsection{Destroy random} 

This method randomly selects a group of tasks from all assigned tasks. The selected tasks are removed from their assigned agents and re-assigned using RMCA.

\subsubsection{Destroy worst}

This strategy randomly \updated{selects} a group of tasks from the agent with the worst TTD. The algorithm \updated{records} the tasks that are selected in a tabu list 
to \updated{avoid selecting them again}. After all tasks are selected once, we clear the tabu list and allow all tasks to be selected again. 

\subsubsection{Destroy multiple}

This method selects a group of agents that have the worst sum of TTD. Then it randomly destroys one task from each agent. It also makes use of a tabu list as in the previous strategy.

\section{Experiments}

\begin{figure}[t]
\centering
\includegraphics[width=0.4\columnwidth]{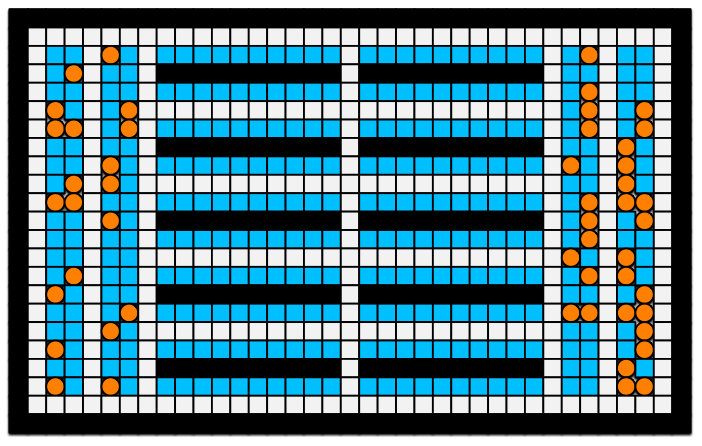}

\caption{A warehouse map with 21 x 35 tiles, where blue tiles are endpoints for tasks, orange tiles are initial locations of the robots, and black tiles are static obstacles.}
\label{figure:warehouse}
\end{figure}

We perform our experiments on a $21 \times 35$ warehouse map as shown in Fig.  \ref{figure:warehouse}, where black tiles are static obstacles, white tiles are corridors, blue tiles represent potential origins and destinations (endpoints) of the tasks, and orange tiles represent 
starting locations of the robots.

For the experiments, we test the performance of the designed algorithms under different instances. Each instance includes a set of packages/tasks with randomly generated origins and destinations and a fleet of robots/agents, where the origin and destination for each task are different\footnote{Our implementation codes of the designed algorithms are available at:\\ https://github.com/nobodyczcz/MCA-RMCA.git}.


\subsection{One-shot Experiment}

\updated{
We first evaluate the designed algorithms in an offline manner to test their scalability. Here, we assume that all the tasks are initially released. This helps us to learn how the number of tasks and other parameters influence the algorithms' performance, and how many tasks our algorithm can process in one assignment time instant.}
\subsubsection{Relative TTD and Runtime}
The first experiment compares variants of methods for different numbers of agents 
and different capacities of agents.
We compare two decoupled versions of the algorithms, where we first complete the task assignment before doing any route planning.  In these variants we use optimal path length as the distance metric while performing task assignment.  We consider two variants: decoupled MCA (MCA-pbs) where we simply assign tasks to the agent which will cause the least delay (assuming optimal path length travel), 
and decoupled RMCA  (RMCA(r)-pbs) where we assign the task with maximum relative regret to its first choice.  
The routing phase uses PBS~\cite{PBS} to rapidly find a set of collision-free routes for the agents given the task assignment.
We compare three coupled approaches:
\ignore{For all of these methods the current travel times are used to compute travel cost. 
}
MCA uses greedy task assignment, while 
RMCA instead uses maximum (absolute or relative) regret to determine which task to assign first.
\ignore{
We consider two forms of regret: absolute regret where which is simply the difference in travel delay
for the first and second options, and relative regret which is defined as the travel delay of the second best option divided by the first option.}
For each number of tasks, each number of agents (Agents) and each capacity (Cap), we randomly generate 25 instances. Each task in each instance randomly selects two endpoints (blue tiles in Fig. \ref{figure:warehouse}) as the start and goal locations for the task.

\begin{figure}[t]
\centering
\includegraphics[width=0.9\columnwidth]{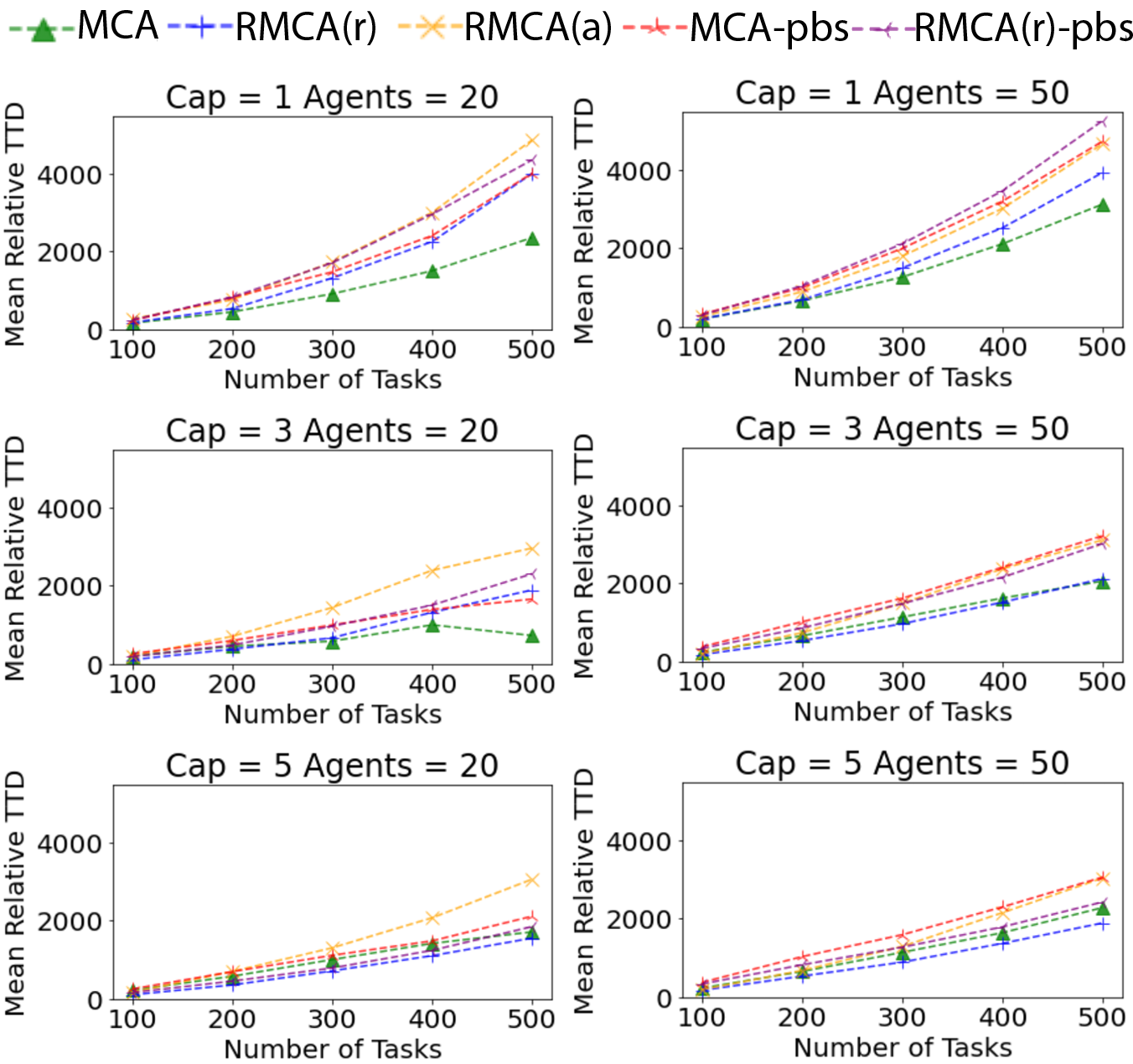}

\caption{Mean relative TTD versus number of tasks on different numbers of agents and different capacity values.}
\label{mca_rmca_delay}
\end{figure}

Fig. \ref{mca_rmca_delay} shows the algorithms' relative TTD. The relative TTD is defined as real TTD minus the TTD of RMCA(r) when ignoring collisions. 
\updated{The reason we use relative TTD as a baseline is that the absolute TTD values in one-shot experiment are very large numbers varying in a relative small range. If using absolute TTD values, it is hard to distinguish the performance difference of algorithms in plots.}
Overall we can see that the decoupled methods are never the best, thus justifying that we want to solve this problem in a coupled manner instead of separate task assignment and routing. For Cap$=1$, MCA is preferable since we cannot modify the route of an agent already assigned to a task to take on a new task and regret is not required.
For Cap$=3$, RMCA(r) eventually becomes the superior approach as the number of agents grows. 
When Cap$=5$, RMCA(r) is clearly the winner. 
Interestingly, the absolute regret based approach RMCA(a) does not perform well at all.
\finalUpdated{This may be because the numbers of tasks assigned to the individual
agents by RMCA(a) are far from even, and the resulting travel delay changes greatly when agents are assigned with more tasks. In other words, RMCA(a) prefers to assign tasks to agents with more tasks.}
\ignore{This may be because the regret comparison, while correct for the current state of agents with their existing tasks, is misleading since the travel delay resulting from RMCA(a) changes greatly when agents with more tasks.} 
The relative regret is more stable to these changes.

\begin{figure}[t]
\centering
\includegraphics[width=0.9\columnwidth]{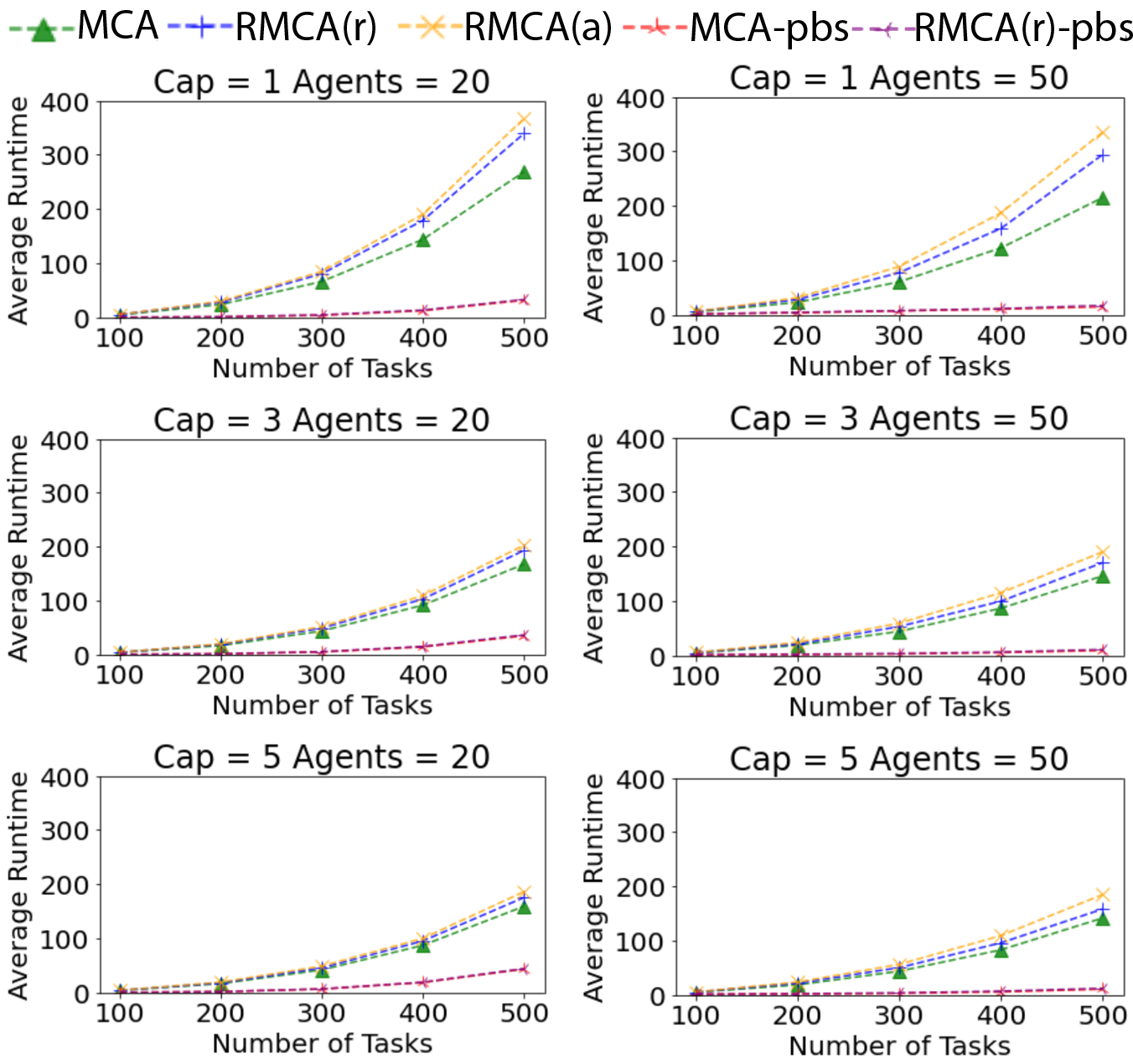}
\caption{Average runtime versus number of tasks on different numbers of agents and different capacity values. }
\label{fig:runtime}
\end{figure}

\begin{table}
\centering
\caption{Mean Relative TTD of Anytime MCA/RMCA on 500 tasks.}
\setlength{\tabcolsep}{2pt}
\resizebox{0.85\columnwidth}{!}{
	\begin{tabular}{|c|c|c|c|c|c|c|c|c|c|c|c|}
			\hline
	        \multirow{3}{*}{Cap}  & \multirow{3}{*}{Agents} & \multirow{3}{*}{RMCA(r)} & \multicolumn{3}{c|}{ RMCA(r)+DR} &
	        \multicolumn{3}{c|}{ RMCA(r)+DW} & \multicolumn{3}{c|}{ RMCA(r)+DM}
	        \\
	        \cline{4-12}
             &&& \multicolumn{3}{c|}{Group Size} & \multicolumn{3}{c|}{Group Size} & \multicolumn{3}{c|}{Group Size} \\
			 \cline{4-12}
             &&& 1 & 3 & 5 & 1 & 3 & 5 & 1 & 3 & 5  \\
			\hline
			1 & 20 & 2762 & 1800 & \textbf{1687} & 1752 & 2108 & 2025 & 2088 & 2714 & 2565 & 2454 \\
			 & 30 & 2871 & 2009 & \textbf{1902} & 1915 & 2276 & 2215 & 2363 & 2827 & 2743 & 2652 \\
			 & 40 & 2876 & 2089 & \textbf{2031} & 2060 & 2367 & 2328 & 2471 & 2836 & 2788 & 2701 \\
			 & 50 & 2906 & 2195 & 2173 & \textbf{2199} & 2481 & 2469 & 2604 & 2887 & 2830 & 2791 \\
			\hline
			3 & 20 & 1085 & 529 & 487 & 470 & 530 & \textbf{416} & 464 & 1058 & 980 & 861 \\
			 & 30 & 1132 & 765 & 710 & 689 & 729 & \textbf{654} & 686 & 1116 & 1074 & 1023 \\
			 & 40 & 1155 & 819 & 798 & \textbf{781} & 812 & 791 & 792 & 1148 & 1129 & 1108 \\
			 & 50 & 1193 & 888 & \textbf{856} & 858 & 875 & 862 & 877 & 1187 & 1171 & 1131 \\
			\hline
			5 & 20 & 726 & 370 & 331 & 319 & 311 & \textbf{253} & 260 & 698 & 635 & 585 \\
			 & 30 & 757 & 452 & 441 & \textbf{415} & 451 & 420 & 433 & 747 & 718 & 687 \\
			 & 40 & 848 & 536 & 511 & 525 & 511 & 482 & \textbf{480} & 839 & 810 & 782 \\
			 & 50 & 906 & 617 & 623 & 623 & 614 & \textbf{574} & 584 & 899 & 883 & 861 \\
			\hline
	        \multirow{3}{*}{Cap}  & \multirow{3}{*}{Agents}& \multirow{3}{*}{MCA} & \multicolumn{3}{c|}{MCA+DR} &
	        \multicolumn{3}{c|}{MCA+DW} & \multicolumn{3}{c|}{MCA+DM}
	        \\
	        \cline{4-12}
             &&& \multicolumn{3}{c|}{Group Size} & \multicolumn{3}{c|}{Group Size} & \multicolumn{3}{c|}{Group Size} \\
			 \cline{4-12}
             &&& 1 & 3 & 5 & 1 & 3 & 5 & 1 & 3 & 5  \\
			\hline
			1 & 20 & 1497 & 850 & 723 & \textbf{715} & 977 & 952 & 976 & 1451 & 1316 & 1252 \\
			 & 30 & 1514 & 927 & 880 & \textbf{873} & 1115 & 1067 & 1138 & 1486 & 1449 & 1412 \\
			 & 40 & 1994 & 1432 & 1406 & \textbf{1376} & 1618 & 1581 & 1696 & 1976 & 1943 & 1908 \\
			 & 50 & 1983 & 1498 & \textbf{1469} & 1480 & 1675 & 1672 & 1769 & 1973 & 1947 & 1915 \\
			\hline
			3 & 20 & 117 & -360 & -395 & -396 & -378 & \textbf{-434} & -428 & 94 & 58 & -21 \\
			 & 30 & 924 & 549 & 510 & 510 & 535 & \textbf{501} & 516 & 913 & 890 & 858 \\
			 & 40 & 1261 & 898 & 868 & \textbf{854} & 879 & 876 & 885 & 1249 & 1227 & 1199 \\
			 & 50 & 1273 & 938 & 925 & \textbf{914} & 931 & 940 & 947 & 1266 & 1245 & 1222 \\
			\hline
			5 & 20 & 748 & 374 & 357 & 337 & 298 & \textbf{276} & 286 & 734 & 689 & 607 \\
			 & 30 & 1197 & 809 & 793 & 778 & 742 & 724 & \textbf{722} & 1178 & 1128 & 1082 \\
			 & 40 & 1367 & 958 & 937 & 966 & 932 & \textbf{899} & 932 & 1347 & 1311 & 1258 \\
			 & 50 & 1266 & 922 & 896 & 915 & 888 & 889 & \textbf{877} & 1258 & 1230 & 1208 \\
			\hline
		\end{tabular}
		}
	\label{table:anytime_mca}
\end{table}

Fig. \ref{fig:runtime} shows the average runtime for the above experiment. The results show that decoupled approaches are advantageous in runtime, especially for instances with a large number of tasks and small capacity. Although RMCA and MCA require more runtime than the decoupled approaches, we demonstrate below that MCA and RMCA are still competitive in runtime compared with other algorithms.

\subsubsection{Anytime Improvement Methods}
\label{sec:anytime-exp}

The second experiment uses any time improvement algorithm to improve the solution from RMCA(r) for $60$ seconds with three neighbourhood destroy strategies:
\finalUpdated{\emph{Destroy random (DR)}, \emph{Destroy worst (DW)} and \emph{Destroy multiple (DM)}}. For each destroy strategy, we run experiments on different destroy group sizes (how many tasks to destroy each time).  The experiment is performed on $25$ instances that each have $500$ tasks with different capacity values and agents' numbers.

\updated{\Cref{table:anytime_mca} shows the results of relative TTD of RMCA(r)/MCA (Relative to the TTD of RMCA(r) that ignores collisions, and the lower the better) under different anytime improvement strategies. The results show that all of the three neighbourhood destroy methods improve the solution quality of RMCA(r) and MCA. We still see that MCA performs better than RMCA(r) when capacity and number of agents are low (The relative TTD of MCA smaller than $0$ means its TTD is smaller than TTD of RMCA(r) that ignores collisions.), even the anytime improvement strategies can not reverse this trend. Overall, destroy random and destroy worst performs better than destroy multiple.
This is not unexpected as simple random neighbourhoods are often very competitive for large neighbourhood search. 
}

\updated{
\subsection{Lifelong Experiment}
}
In this part, we test the performance of RMCA(r) in a lifelong setting compared with the TPTS and CENTRAL algorithms in  \cite{ma2017lifelong}. The MAPD problem solved by TPTS and CENTRAL
assumes that each agent can carry a maximum of one package at a time, and the objective is to minimize the makespan. 
This objective is somewhat misleading when we
consider the continuous nature of the underlying problem
where new tasks arrive as the plan progresses. As a result, minimizing TTD might be a better objective since it may help in optimizing the total throughput of the system by trying to make agents idle as soon as possible, whereas with makespan
minimization all agents can be active until the last time point.

At each timestep, after adding newly released tasks to the unassigned task set $\mathcal P^u$, the system performs RMCA(r) on current assignments set $\mathcal A$, and runs the anytime improvement process on all released tasks that are not yet picked up. The RMCA(r) uses the anytime improvement strategy of destroy random with a group size of $5$. As the anytime improvement triggers at every timestep when new tasks arrive, and involves all released yet unpicked up tasks, we set the improvement time as $1$ second in each run.

We generate $25$ instances with $500$ tasks. For each instance, we use different task release frequencies ($f$): $0.2$ (release $1$ task every $5$ timestep), $2$ and $10$ ($10$ tasks are released each timestep). For each task release frequency, we test the performance of the algorithms under different agent capacities (Cap) and different numbers of agents (Agents).


\begin{table}
\centering
	\caption{Lifelong Experiment on different algorithms.}
	\setlength{\tabcolsep}{0.5pt}
\resizebox{0.9\columnwidth}{!}{
		\begin{tabular}{|c|c|c|c|c|c|c|c|c|c|c|c|c|c|c|c|}
			\hline
	        \multirow{2}{*}{$f$} & \multirow{2}{*}{Cap}  & \multirow{2}{*}{Agents}&
	        \multicolumn{3}{c|}{RMCA(r) Anytime} &
	        \multicolumn{3}{c|}{CENTRAL} &
	        \multicolumn{3}{c|}{TPTS} \\
	        \cline{4-12}
	        
	        & & & TTD & Makespan &  T/TS & TTD & Makespan &  T/TS & TTD & Makespan &  T/TS \\
	        
			\hline
			0.2 & 1 & 20 & \textbf{3138} & \textbf{2526} & 0.205 & 4365 & 2528 & 0.364 & 3645 & 2528 & \textbf{0.103} \\
			 &  & 30 & \textbf{2729} & \textbf{2525} & \textbf{0.208} & 3864 & 2527 & 0.762 & 3002 & 2526 & 0.242 \\
			 &  & 40 & \textbf{2297} & \textbf{2523} & \textbf{0.210} & 3572 & 2527 & 1.300 & 2646 & 2525 & 0.442 \\
			 &  & 50 & \textbf{2176} & \textbf{2523} & \textbf{0.214} & 3394 & 2525 & 1.945 & 2456 & 2524 & 0.710 \\
			\cline{2-12}
			 & 3 & 20 & 3056 & 2526 & 0.207 & -- & -- & -- & -- & -- & -- \\
			 & & 30 & 2661 & 2525 & 0.210 & -- & -- & -- & -- & -- & -- \\
			 & & 40 & 2223 & 2523 & 0.216 & -- & -- & -- & -- & -- & -- \\
			& & 50 & 2121 & 2523 & 0.219 & -- & -- & -- & -- & -- & -- \\
			\cline{2-12}
			 & 5 & 20 & 3056 & 2526 & 0.207 & -- & -- & -- & -- & -- & -- \\
			&  & 30 & 2661 & 2525 & 0.211 & -- & -- & -- & -- & -- & -- \\
		     &  & 40 & 2223 & 2523 & 0.217 & -- & -- & -- & -- & -- & -- \\
		     &  & 50 & 2121 & 2523 & 0.219 & -- & -- & -- & -- & -- & -- \\
			\hline
			2 & 1 & 20 & \textbf{65938} & 626 & 0.489 & 75294 & \textbf{610} & 0.125 & 82734 & 639 & \textbf{0.022} \\
			&  & 30 & \textbf{30317} & \textbf{436} & 0.705 & 37327 & 446 & 0.284 & 47252 & 490 & \textbf{0.099} \\
		    &  & 40 & \textbf{13945} & \textbf{344} & 0.884 & 19930 & 376 & 0.426 & 30491 & 413 & \textbf{0.273} \\
			&  & 50 & \textbf{6279} & \textbf{300} & 1.022 & 11185 & 328 & \textbf{0.615} & 21853 & 377 & 0.660 \\
			\cline{2-12}
			 & 3 & 20 & 17904 & 349 & 0.791 & -- & -- & -- & -- & -- & -- \\
			 &  & 30 & 7504 & 302 & 0.933 & -- & -- & -- & -- & -- & -- \\
			&  & 40 & 4644 & 291 & 0.999 & -- & -- & -- & -- & -- & -- \\
			&  & 50 & 3475 & 290 & 1.045 & -- & -- & -- & -- & -- & -- \\
			\cline{2-12}
			 & 5 & 20 & 12711 & 320 & 0.860 & -- & -- & -- & -- & -- & -- \\
			 &  & 30 & 7005 & 299 & 0.942 & -- & -- & -- & -- & -- & -- \\
			 &  & 40 & 4670 & 291 & 1.002 & -- & -- & -- & -- & -- & -- \\
			 &  & 50 & 3463 & 288 & 1.053 & -- & -- & -- & -- & -- & -- \\
			\hline
			10 & 1 & 20 & \textbf{106290} & 624 & 0.142 & 116357 & \textbf{587} & 0.421 & 125374 & 626 & \textbf{0.025} \\
			 &  & 30 & \textbf{68166} & 435 & 0.210 & 76934 & \textbf{419} & 1.062 & 86267 & 462 & \textbf{0.086} \\
			 &  & 40 & \textbf{49140} & 338 & 0.284 & 56896 & \textbf{337} & 2.426 & 66171 & 383 & \textbf{0.238} \\
			 &  & 50 & \textbf{38050} & \textbf{280} & \textbf{0.362} & 45170 & 288 & 2.828 & 55409 & 339 & 0.559 \\
			\cline{2-12}
			 & 3 & 20 & 52771 & 322 & 0.209 & -- & -- & -- & -- & -- & -- \\
			 &  & 30 & 31832 & 226 & 0.305 & -- & -- & -- & -- & -- & -- \\
			 &  & 40 & 21651 & 179 & 0.404 & -- & -- & -- & -- & -- & -- \\
			 &  & 50 & 15851 & 153 & 0.521 & -- & -- & -- & -- & -- & -- \\
			\cline{2-12}
			 & 5 & 20 & 36790 & 247 & 0.271 & -- & -- & -- & -- & -- & -- \\
			 &  & 30 & 21723 & 176 & 0.384 & -- & -- & -- & -- & -- & -- \\
			 &  & 40 & 14970 & 145 & 0.496 & -- & -- & -- & -- & -- & -- \\
			 &  & 50 & 11464 & 129 & 0.613 & -- & -- & -- & -- & -- & -- \\
			\hline
		\end{tabular}
		}
	\label{table:lifelong}

\end{table}
 
\begin{table}
	\begin{center}
	\caption{t-test compares RMCA(r) to CENTRAL and TPTS}
	\label{table:t_test}
    \resizebox{0.8\columnwidth}{!}{
		\begin{tabular}{|c|c|c|c|c|}
			\hline
	          & \multicolumn{2}{c|}{CENTRAL} & \multicolumn{2}{c|}{TPTS} \\
	         \hline
	          & TTD & Makespan & TTD & Makespan \\
			\hline
	         t-score & -17.01 & -0.06 & -22.43 & 1.83 \\
			\hline
	         p-value & $3.47\times 10^{-53}$ & $ 0.95 $ & $ 2.89 \times 10^{-81}$ & $0.06$\\
			\hline
		\end{tabular}
		}
		\end{center}

\end{table}

\subsubsection{Result}
\Cref{table:lifelong} shows that RMCA(r) not only optimizes TTD, its makespans are overall close to CENTRAL, and are much better than TPTS. Comparing TTD, CENTRAL and TPTS perform much worse than RMCA(r). This supports our argument that makespan is not sufficient for optimizing the total throughput of the system. In addition, the \finalUpdated{runtime per timestep (T/TS)} shows that RMCA(r) gets a better solution quality while consuming less runtime on each timestep compared with CENTRAL. A lower runtime per timestep makes RMCA(r) better suited to real-time lifelong operations. Furthermore, by increasing the capacity of robots, both total travel delay and makespan are reduced significantly, which increases the throughput and efficiency of the warehouse.

\subsubsection{T-Test on TTD and Makespan}
We evaluate how significant is the solution quality of RMCA(r) with respect to CENTRAL and TPTS by performing t-test with significance level of $0.1$ on the normalized TTD and normalized makespan for experiments with robots' Cap$=1$. 
The \emph{normalized TTD} is defined as $\frac{TTD \cdot N_a}{N_t \cdot f}$
where $N_t$ is the number of tasks, $N_a$ is the number of agents and $f$ is the task frequency. This definition is based on the observation that increasing $N_a$ decreases TTD, and increasing $N_t$ and $f$ increases TTD. Similarly \emph{normalized makespan} is $\frac{makespan \cdot N_a \cdot f}{N_t}$
(where now increasing $f$ decreases makespan). 
\Cref{table:t_test} shows the $t$-score and $p$-value for the null hypotheses that RMCA(r) and the other methods are identical. The results show that RMCA(r) significantly improves the normalized TTD compared with CENTRAL and TPTS and improves the normalized makespan compared with TPTS.

\section{Conclusion}

In this paper, we have designed two algorithms MCA and RMCA to solve the Multi-agent Pickup and Delivery problem where each robot
can carry multiple packages simultaneously. MCA and RMCA successfully perform task assignment and path planning simultaneously. 
This is achieved by \updated{using the real collision-free costs to guide the multi-task multi-robot assignment process}. Further, we observe that the newly introduced anytime improvement strategy improves solutions substantially.
\finalUpdated{Future work will extend the anytime improvement strategies to refine the agents' routes, and improve the algorithms' completeness on path planning.}

\ignore{
The future research will focuses on the online experiments and improving the anytime improvement strategy to include the refining of path planning result, not only the refining of task assignment result. By doing this we will able to further improve the solution quality.
}

\bibliographystyle{IEEEtran}

\bibliography{Reference210112Revise}

\begin{thebibliography}{10}
\providecommand{\url}[1]{#1}
\csname url@rmstyle\endcsname
\providecommand{\newblock}{\relax}
\providecommand{\bibinfo}[2]{#2}
\providecommand\BIBentrySTDinterwordspacing{\spaceskip=0pt\relax}
\providecommand\BIBentryALTinterwordstretchfactor{4}
\providecommand\BIBentryALTinterwordspacing{\spaceskip=\fontdimen2\font plus
\BIBentryALTinterwordstretchfactor\fontdimen3\font minus
  \fontdimen4\font\relax}
\providecommand\BIBforeignlanguage[2]{{%
\expandafter\ifx\csname l@#1\endcsname\relax
\typeout{** WARNING: IEEEtran.bst: No hyphenation pattern has been}%
\typeout{** loaded for the language `#1'. Using the pattern for}%
\typeout{** the default language instead.}%
\else
\language=\csname l@#1\endcsname
\fi
#2}}

\bibitem{ma2017lifelong}
H.~Ma, J.~Li, T.~Kumar, and S.~Koenig, ``Lifelong multi-agent path finding for
  online pickup and delivery tasks,'' in \emph{AAMAS}, 2017, pp. 837--845.

\bibitem{korsah2013comprehensive}
G.~A. Korsah, A.~Stentz, and M.~B. Dias, ``A comprehensive taxonomy for
  multi-robot task allocation,'' \emph{The International Journal of Robotics
  Research}, vol.~32, no.~12, pp. 1495--1512, 2013.

\bibitem{SternSoCS19}
R.~Stern, N.~R. Sturtevant, A.~Felner, S.~Koenig, H.~Ma, T.~T. Walker, J.~Li,
  D.~Atzmon, L.~Cohen, T.~K.~S. Kumar, R.~Bart{\'{a}}k, and E.~Boyarski,
  ``Multi-agent pathfinding: Definitions, variants, and benchmarks,'' in
  \emph{Proceedings of the International Symposium on Combinatorial Search
  (SoCS)}, 2019, pp. 151--159.

\bibitem{honig2018conflict}
W.~H{\"o}nig, S.~Kiesel, A.~Tinka, J.~Durham, and N.~Ayanian, ``Conflict-based
  search with optimal task assignment,'' in \emph{Proceedings of the
  International Joint Conference on Autonomous Agents and Multiagent Systems},
  2018, pp. 757--765.

\bibitem{liu2019task}
M.~Liu, H.~Ma, J.~Li, and S.~Koenig, ``Task and path planning for multi-agent
  pickup and delivery.'' in \emph{AAMAS}, 2019, pp. 1152--1160.

\bibitem{hai_robots}
\BIBentryALTinterwordspacing
``Haipick system.'' [Online]. Available:
  \url{https://www.hairobotics.com/product/index}
\BIBentrySTDinterwordspacing

\bibitem{meng2019idle}
N.~Meng~Kou, C.~Peng, H.~Ma, T.~Satish~Kumar, and S.~Koenig, ``Idle time
  optimization for target assignment and path finding in sortation centers,''
  \emph{arXiv}, pp. arXiv--1912, 2019.

\bibitem{silver2005cooperative}
D.~Silver, ``Cooperative pathfinding.'' \emph{AIIDE}, vol.~1, pp. 117--122,
  2005.

\bibitem{LNS}
P.~Shaw, ``Using constraint programming and local search methods to solve
  vehicle routing problems,'' in \emph{CP~1998}, 1998, pp. 417--431.

\bibitem{nguyen2019generalized}
V.~Nguyen, P.~Obermeier, T.~C. Son, T.~Schaub, and W.~Yeoh, ``Generalized
  target assignment and path finding using answer set programming,'' in
  \emph{IJCAI}, 2019, pp. 1216--1223.

\bibitem{bai2020event}
X.~Bai, M.~Cao, and W.~Yan, ``Event- and time-triggered dynamic task
  assignments for multiple vehicles,'' \emph{Autonomous Robots}, vol.~44,
  no.~5, pp. 877--888, 2020.

\bibitem{toth2014vehicle}
P.~Toth and D.~Vigo, \emph{Vehicle routing: problems, methods, and
  applications}.\hskip 1em plus 0.5em minus 0.4em\relax Siam, 2014, vol.~18.

\bibitem{bai2018integrated}
X.~Bai, W.~Yan, S.~S. Ge, and M.~Cao, ``An integrated multi-population genetic
  algorithm for multi-vehicle task assignment in a drift field,''
  \emph{Information Sciences}, vol. 453, pp. 227--238, 2018.

\bibitem{bai2019efficient}
X.~Bai, M.~Cao, W.~Yan, and S.~S. Ge, ``Efficient routing for
  precedence-constrained package delivery for heterogeneous vehicles,''
  \emph{IEEE Transactions on Automation Science and Engineering}, vol.~17,
  no.~1, pp. 248--260, 2019.

\bibitem{cordeau2007dial}
J.-F. Cordeau and G.~Laporte, ``The dial-a-ride problem: models and
  algorithms,'' \emph{Annals of Operations Research}, vol. 153, no.~1, pp.
  29--46, 2007.

\bibitem{tillman1972upperbound}
F.~A. Tillman and T.~M. Cain, ``An upperbound algorithm for the single and
  multiple terminal delivery problem,'' \emph{Management Science}, vol.~18,
  no.~11, pp. 664--682, 1972.

\bibitem{potvin1993parallel}
J.-Y. Potvin and J.-M. Rousseau, ``A parallel route building algorithm for the
  vehicle routing and scheduling problem with time windows,'' \emph{European
  Journal of Operational Research}, vol.~66, no.~3, pp. 331--340, 1993.

\bibitem{koenig2008agent}
S.~Koenig, X.~Zheng, C.~A. Tovey, R.~B. Borie, P.~Kilby, V.~Markakis,
  P.~Keskinocak, \emph{et~al.}, ``Agent coordination with regret clearing.'' in
  \emph{AAAI}, 2008, pp. 101--107.

\bibitem{sharon2015conflict}
G.~Sharon, R.~Stern, A.~Felner, and N.~R. Sturtevant, ``Conflict-based search
  for optimal multi-agent pathfinding,'' \emph{Artificial Intelligence}, vol.
  219, pp. 40--66, 2015.

\bibitem{bettinelli2011branch}
A.~Bettinelli, A.~Ceselli, and G.~Righini, ``A branch-and-cut-and-price
  algorithm for the multi-depot heterogeneous vehicle routing problem with time
  windows,'' \emph{Transportation Research Part C: Emerging Technologies},
  vol.~19, no.~5, pp. 723--740, 2011.

\bibitem{EPEA}
M.~Goldenberg, A.~Felner, R.~Stern, G.~Sharon, N.~R. Sturtevant, R.~C. Holte,
  and J.~Schaeffer, ``Enhanced partial expansion {A},'' \emph{Journal of
  Artificial Intelligence Research}, vol.~50, pp. 141--187, 2014.

\bibitem{SurynekIJCAI19}
P.~Surynek, ``Unifying search-based and compilation-based approaches to
  multi-agent path finding through satisfiability modulo theories,'' in
  \emph{IJCAI}, 2019, pp. 1177--1183.

\bibitem{CohenIJCAI16}
L.~Cohen, T.~Uras, T.~K.~S. Kumar, H.~Xu, N.~Ayanian, and S.~Koenig, ``Improved
  solvers for bounded-suboptimal multi-agent path finding,'' in \emph{IJCAI},
  2016, pp. 3067--3074.

\bibitem{PBS}
H.~Ma, D.~Harabor, P.~J. Stuckey, J.~Li, and S.~Koenig, ``Searching with
  consistent prioritization for multi-agent path finding,'' in \emph{AAAI},
  2019, pp. 7643--7650.

\bibitem{PLPushAndSwap}
Q.~Sajid, R.~Luna, and K.~E. Bekris, ``Multi-agent pathfinding with
  simultaneous execution of single-agent primitives,'' in \emph{SoCS}, 2012,
  pp. 88--96.

\bibitem{li2020lifelong}
J.~Li, A.~Tinka, S.~Kiesel, J.~W. Durham, T.~Kumar, and S.~Koenig, ``Lifelong
  multi-agent path finding in large-scale warehouses,'' \emph{arXiv preprint
  arXiv:2005.07371}, 2020.

\bibitem{honig2019}
W.~H{\"o}nig, S.~Kiesel, A.~Tinka, J.~W. Durham, and N.~Ayanian, ``Persistent
  and robust execution of {MAPF} schedules in warehouses,'' \emph{IEEE Robotics
  and Automation Letters}, vol.~4, no.~2, pp. 1125--1131, 2019.

\bibitem{ma2017multi}
H.~Ma, T.~S. Kumar, and S.~Koenig, ``Multi-agent path finding with delay
  probabilities,'' in \emph{AAAI}, 2017, pp. 3605--3612.

\bibitem{atzmon}
D.~Atzmon, R.~Stern, A.~Felner, G.~Wagner, R.~Bart{\'a}k, and N.-F. Zhou,
  ``Robust multi-agent path finding,'' in \emph{SOCS}, 2018, pp. 1862--1864.

\bibitem{honig2016multi}
W.~H{\"o}nig, T.~S. Kumar, L.~Cohen, H.~Ma, H.~Xu, N.~Ayanian, and S.~Koenig,
  ``Multi-agent path finding with kinematic constraints.'' in \emph{ICAPS},
  vol.~16, 2016, pp. 477--485.

\end{thebibliography}

\end{document}